\documentclass[pra,twocolumn,aps,letterpaper, preprintnumbers,superscriptaddress]{revtex4-2}
\usepackage{graphicx}
\usepackage{dcolumn}
\usepackage{bm}
\usepackage{adjustbox}
\usepackage{amsthm,amsmath,amssymb}
\usepackage{mathrsfs}
\usepackage{appendix}
\usepackage{amstext} 
\usepackage{url}
\usepackage{float} 
\usepackage[usenames,dvipsnames]{color}
\usepackage[colorlinks=true,citecolor=blue,urlcolor=black]{hyperref}
\usepackage{dcolumn}
\usepackage{booktabs}
\usepackage{graphicx}
\usepackage[normalem]{ulem}
\usepackage[linesnumbered,algoruled,boxed,lined]{algorithm2e} 
\usepackage{siunitx}
\usepackage{makecell}
\usepackage{soul}

\begin{document}
	
	\title{Multi-node quantum key distribution network using existing underground optical fibre infrastructure}
	
	\author{Mariella Minder}
	\email{Corresponding author: mariella.minder@cut.ac.cy}
	\affiliation{Department of Electrical Engineering, Computer Engineering and Informatics, Cyprus University of Technology, Limassol 3036, Cyprus}
	\author{Andreas Siakolas}
	\author{Stephanos Yerolatsitis}
	\author{Konstantinos Katzis}
	\author{Kyriacos Kalli}
	
	\affiliation{Department of Electrical Engineering, Computer Engineering and Informatics, Cyprus University of Technology, Limassol 3036, Cyprus} 
	
	\date{May 29, 2025}
	
	\begin{abstract}
	Quantum key distribution (QKD) offers unconditional information security by allowing two distant users to establish a common encryption key resilient to hacking. 
	Resultingly, QKD networks interconnecting critical infrastructure and enabling the secure exchange of classified information, can provide a solution to the increasing number of successful cyberattacks. 
	To efficiently deploy quantum networks, the technology must be integrated over existing communication infrastructure, such as optical fibre links. 
	Yet, QKD poses stringent requirements on the conditions of the network over which it is deployed. 
	This work demonstrates the first quantum communication network in Cyprus via the deployment of a multi-node quantum network, exploiting existing commercial underground optical fibre. 
	The network employs bidirectional occupation of fibres and wavelength multiplexing in a ring architecture to achieve, with minimal use of dark fibres, high-rate QKD. 
	Results obtained reveal consistent key generation rates across all nodes, confirming reliable operation in a real-world environment. 
	This deployment highlights the feasibility of leveraging existing telecom infrastructure for quantum-secured communication, marking a significant step toward scalable and cost-effective quantum networks suited for critical applications.
	\end{abstract} 
	
	%
	
	\maketitle
	
	\section{Introduction}
	With the radical growth of digital communication needs, the threat of cyberattacks on sensitive information has risen.
	Conventional encryption systems base their security on computational complexity and are therefore fundamentally vulnerable to attackers with computational power beyond the assumed limit, particularly quantum computers. 
	Instead, quantum key distribution (QKD) leverages the principles of quantum mechanics to guarantee that eavesdropping attempts will be detected as noise in the measured quantum states~\cite{Bennett2014}. 
	This achieves information-theoretic security, a level of protection unattainable by classical cryptographic techniques.
	
	Similar to classical communications, to enable secure communication at scale, QKD must be deployed over networked architectures, facilitating key exchange between multiple distant users. 
	However, integrating QKD into existing infrastructure presents significant challenges. 
	The quantum states required to perform QKD are generated by encoding information on properties of single photons, as opposed to conventional communications where bright light is used. 
	Consequently, quantum channels connecting end-users must be low-loss and low noise and, in deployments, dark fibre is often used to meet these stringent requirements. 
	This threatens the practical integration of the technology into current infrastructure and its scalability potential due to the limited availability of dark fibre within existing networks. 
	
	At the same time, a multi-user QKD network deployed over real-life infrastructure must adapt to the fibre architecture available. 
	Existing fibre networks, designed for classical data transmission, are complex and layered. 
	End-users are interconnected via distribution nodes with multiple communication links sharing common optical paths.
	These environments are a hurdle for the integration of QKD and may require trade-offs in functionality, performance and latency, or the use of tools with limited commercial readiness such as optical switching and wavelength multiplexing. 
	Overcoming these barriers without compromising the security of QKD is crucial for the timely adoption of quantum communications.
	Significant progress has been made in the development of quantum communication networks, with various demonstrations over the past decade. 
	Early experiments focused on point-to-point QKD links over dedicated optical fibres, proving the feasibility of secure key exchange over tens to hundreds of kilometres~\cite{Boaron2018} in dark or live fibre~\cite{Dynes2016}. 
	More recent efforts have expanded toward multi-node networks~\cite{Chen2021}, as well as long-distance deployments using trusted relays~\cite{Chen2021_space} and novel protocols~\cite{Pittaluga2025}. 
	Additionally, researchers have explored the feasibility of quantum networking over fibre channels under special conditions such as undersea fibre~\cite{Amies-King2023, Ribezzo2023}, aerial fibre~\cite{Dixon2015, Tang2023} and even hollow-core fibre~\cite{Minder2023}. 
	Despite these achievements, most demonstrations rely on simple fibre infrastructure limiting the research into real-life integration, scalability and resource-effectiveness. 
	
	Our work builds upon prior research by exploring the integration of QKD into existing dark fibre infrastructure, aiming to demonstrate a practical and resilient approach for real-world quantum-secured networks. 
	Specifically, we present the deployment and performance analysis of a four-node QKD ring network implemented over existing but restricted fibres in a metropolitan area. 
	We utilise bidirectionality and wavelength multiplexing techniques to reduce fibre infrastructure requirements without affecting the resulting performance. 
	The network encompasses all layers required for real-life quantum communication, including a key management service and an application layer. 
	The network’s feasibility and key generation performance are evaluated, demonstrating the potential for integrating QKD within real-world telecommunications networks.

	\section{Network details}
	
	\begin{figure*}[h!]
		\centering 
		\includegraphics[width=1\linewidth]{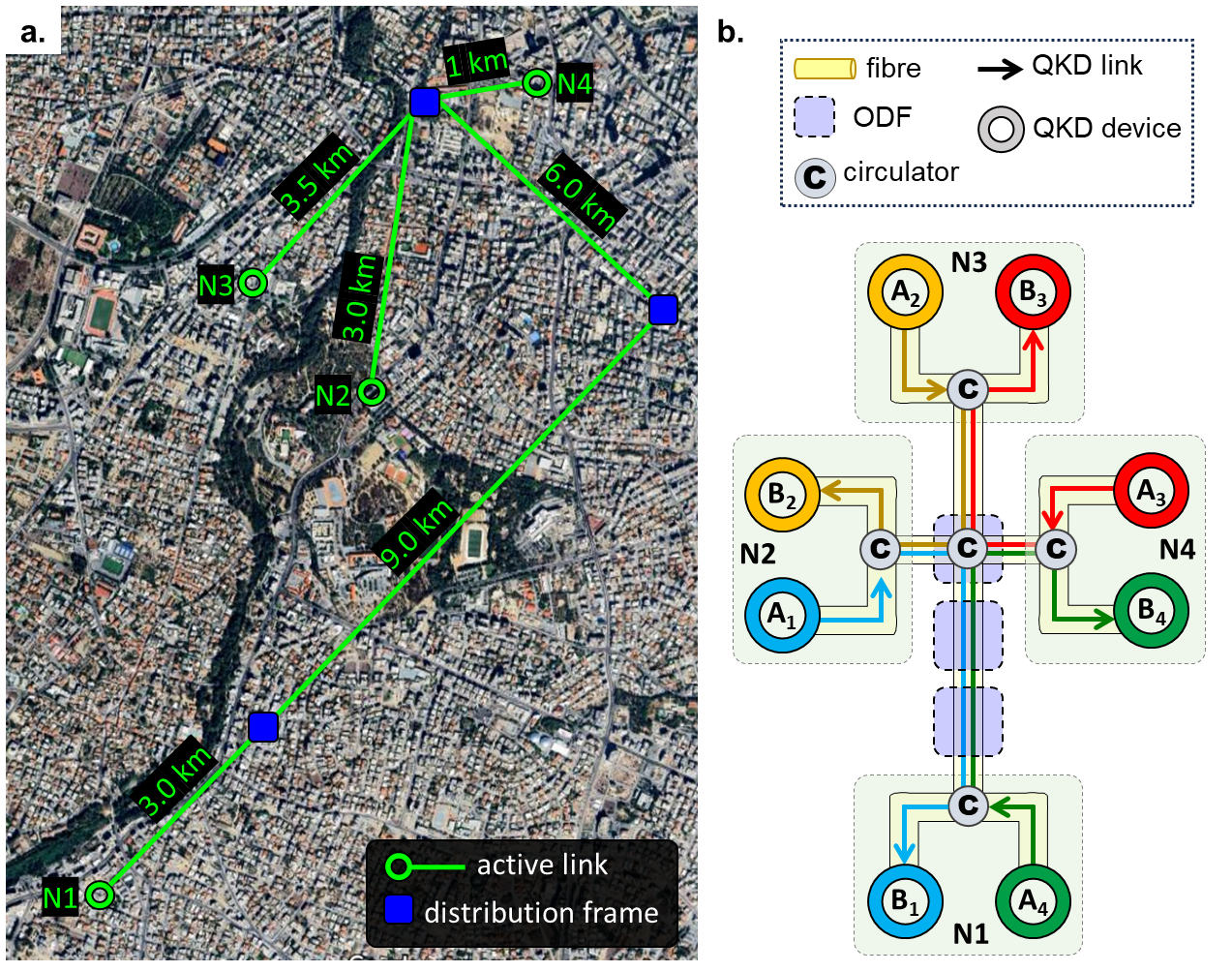}
		\caption{\label{fig:networkDiagram}\textbf{a.} 
			Map of the four-node (N1-N4) quantum network in Nicosia. 
			The available fibre infrastructure interconnects the nodes via optical distribution frames (ODFs). 
			Each link segment (green line) consists of two dark fibres which vary in length. 
			\textbf{b.} 
			Architecture of the quantum channels. 
			A single fibre per segment carries all quantum signals.
			Circulators at the nodes direct the light from the transmitters, A\textsubscript{N}, to the channel and then to the receivers, B\textsubscript{N}. 
			A circulator at an ODF directs the signals from N4-N1.}
	\end{figure*}
	
	In this work, we present the first quantum communication network in Cyprus and the broader Southeast European region, marking a cybersecurity milestone. 
	The network interconnects four governmental authorities in Nicosia over existing underground commercial fibre infrastructure. 
	Originally designed for classical communication, the fibre network connects nodes via optical distribution frames (ODFs), which act as intermediate points between links, Fig.~\ref{fig:networkDiagram}a. 
	Two fibres were available per segment—fibre 1 for quantum signals and fibre 2 for classical traffic. 
	By physically separating quantum and classical signals, noise is reduced, enhancing QKD performance. 
	
\begin{figure*}[h!]
		\centering 
		\includegraphics[width=1\linewidth]{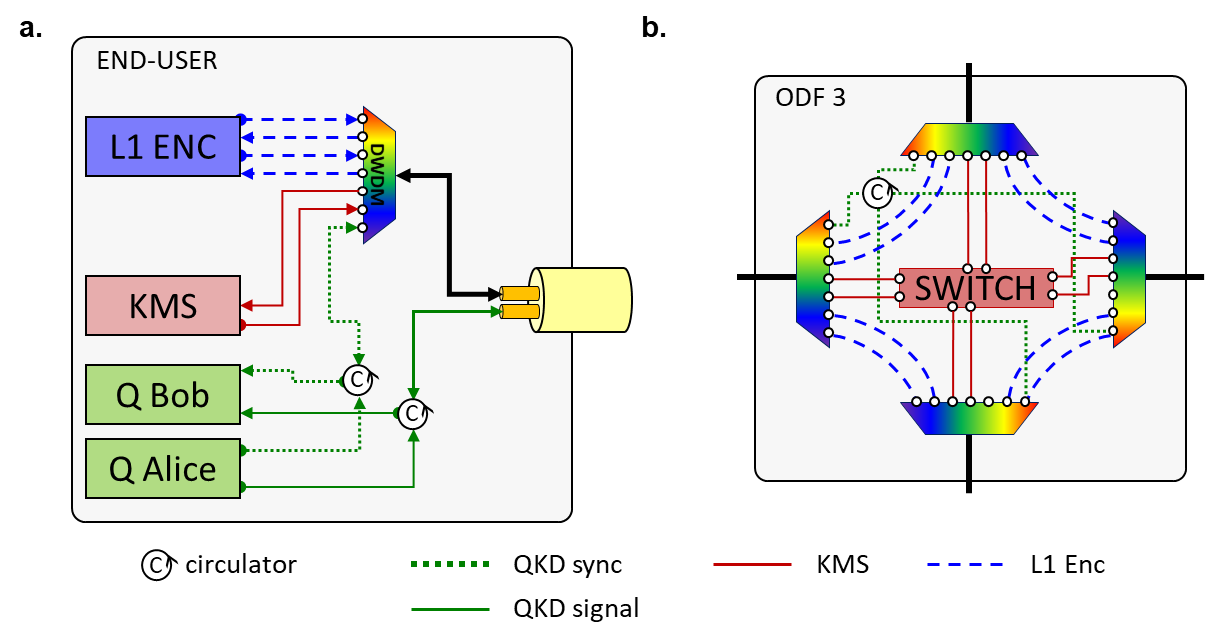}
		\caption{\label{fig:nodeDiagram}
			\textbf{a.} 
			End-user setup: Each node includes two QKD devices, a KMS unit, and a layer~1 encryptor (L1~ENC). 
			Classical signals are multiplexed via DWDM and sent over one fibre; quantum signals use a separate fibre.
			\textbf{b.} 
			ODF 3 design: Encryption and clock signals follow a ring topology using DWDMs and a circulator, while the KMS layer maintains mesh connectivity.}
	\end{figure*}

	A ring topology was implemented at the QKD layer, with each node serving as a trusted node enabling efficient any-to-any key distribution, Fig.~\ref{fig:networkDiagram}b. 
	Each fibre segment supports two QKD links in opposite directions, enabling bidirectional use without wavelength multiplexing. 
	A four-port circulator at ODF~3 enforces anti-clockwise signal flow, while ODFs 1 and 2 remain passive. 
	At the user end, Fig.~\ref{fig:nodeDiagram}a, each node includes a QKD transmitter (Alice), receiver (Bob), and a three-port circulator. 
	The second fibre carries synchronisation, key management (KMS), and application-layer traffic. 
	The application layer uses 10~G layer~1 encryptors in a two-way ring, the sync signal mirrors the quantum ring, and the KMS uses a full-mesh configuration. 
	ODF~3 integrates DWDMs, an optoelectrical switch for mesh relaying, and a circulator for synchronisation handling, Fig.~\ref{fig:nodeDiagram}b. 
	In total, seven ITU-T grid wavelengths are used, one per classical signal.
	
	\section{Results}
	
	\begin{figure*}[h!]
		\centering 
		\includegraphics[width=1\linewidth]{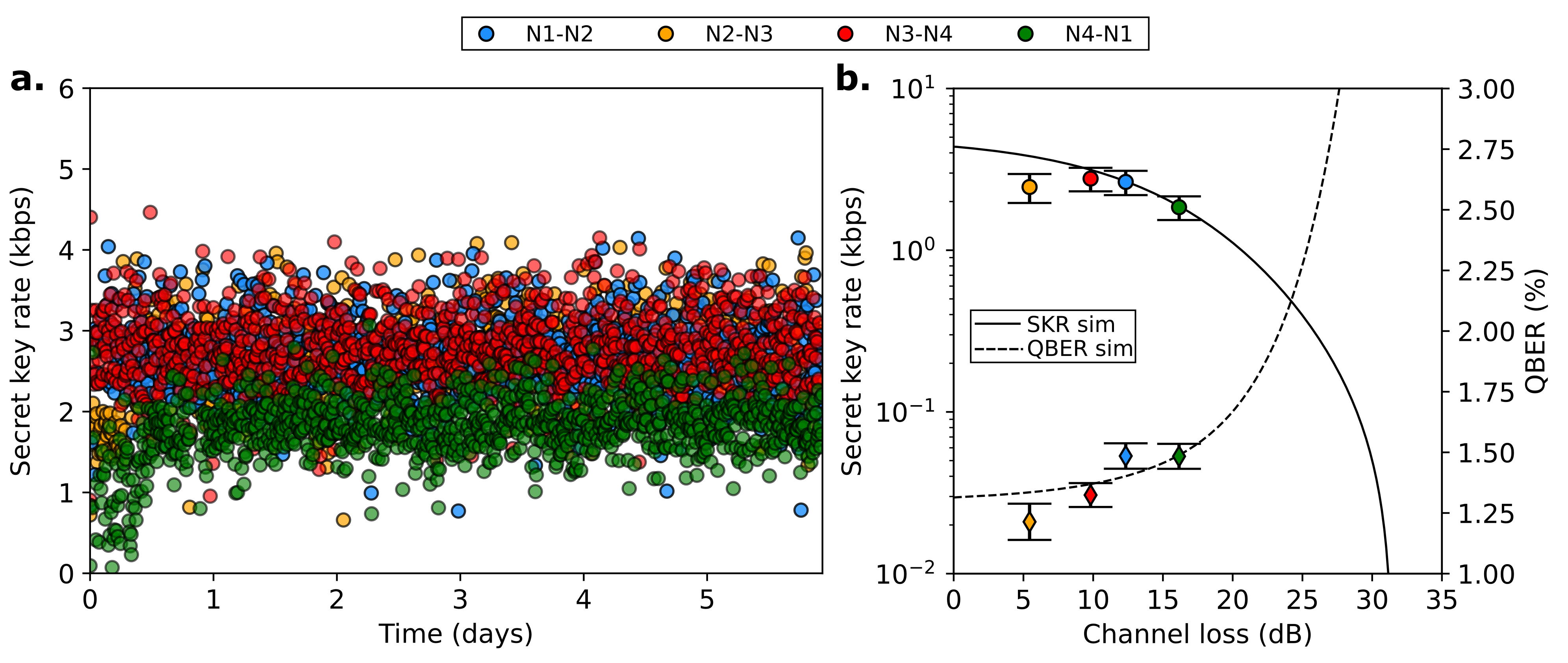}
		\caption{\label{fig:results}
			\textbf{a.} Secret key rates acquired over all links for the duration of 5 days. 
			\textbf{b.} Average secret key rate (circles) and QBER (diamonds) per link compared with the theoretically expected lines (SKR sim, QBER sim).}
	\end{figure*}
	
	The QKD devices deployed in the network, QTI Quell-X~\cite{Francesconi2024}, implement the time-bin encoded, three-state BB84 QKD protocol~\cite{Rusca2018} with weak coherent pulses at a source repetition rate of 600~MHz and a central wavelength of 1545.32~nm. 
	We test the resulting performance of the links in terms of secret key rate (SKR) and quantum bit error rate (QBER) over a period of 5 days. 
	The results are shown in Fig.~\ref{fig:results}. 
	The Alice-Bob pairs achieve an average SKR over all links of $2.4\pm 0.2$~kbps. 
	While the SKR variation is significant, this agrees with the values observed in the pre-deployment in our lab, indicating this is an intrinsic characteristic of the devices and not an effect of the environment over which they are deployed.
	To assess the performance, we plot the average SKR and QBER vs channel loss per link alongside the corresponding theoretical lines by adapting the three-state analysis in~\cite{Rusca2018, Rusca2018finite} to the systems’ architecture. 
	For the simulation, we assume 4\% total detection efficiency, a $40~\mu~s$ average detector deadtime and a dark count probability, p\textsubscript{dc}, of $8.5\times10^{-7}$, consistent with the experimental parameters of the systems. 
	The deployment results align closely with theoretical predictions, confirming that the devices operate near their optimal performance given their intrinsic parameters. 
	Even though the deployed fibres were part of cables carrying live traffic on adjacent fibres, no measurable leakage impacting the QBER was observed, confirming the suitability of integrating quantum communication systems into Cyprus's commercial fibre network.
	
	\section{Conclusion}
	
	This work presents the first deployment of a QKD network in Cyprus and the broader Southeast European region, leveraging existing underground fibre infrastructure. 
	The network interconnects four governmental authorities in a metropolitan setting using a ring architecture with trusted nodes, bidirectional transmission, and wavelength multiplexing to achieve high-performance QKD with minimal infrastructure requirements. 
	By separating quantum and classical signals and carefully adapting to the constraints of a commercial fibre environment, the deployment demonstrates a scalable, resource-efficient, and operationally resilient solution for real-world quantum-secured communication.
	
	Performance evaluation over several days confirms stable secret key generation across all links, with results matching theoretical expectations. 
	The integration of a full quantum layer, key management service, and application-layer encryption illustrates the system’s completeness and readiness for practical use. 
	This demonstration underscores the viability of QKD in realistic telecom environments and marks a step toward broader adoption of quantum networks for securing sensitive information in critical infrastructure. 
	Future work will focus on expanding network scale, exploring dynamic reconfiguration, and integrating quantum-safe technologies alongside QKD to support evolving cybersecurity demands.

	\acknowledgments
	This work is co-funded by the European Commission and the Cyprus Deputy Ministry of Research, Innovation and Digital Policy under the Grant Agreement No. 101091655.
	
	\bibliography{spie3node} 
	\bibliographystyle{spiebib} 

\end{document}